\documentclass[onecolumn,amsmath,amssymb]{snp}
\usepackage{graphicx}
\usepackage{dcolumn}
\usepackage{bm}
\begin{document}

\title{{\Large Effective relativistic mean field model for finite nuclei 
and neutron  stars}} 
\author{\large Bharat Kumar$^{1,3}$}
\email{bharat@iopb.res.in}
\author{\large B. K. Agrawal$^{2,3}$}
\author{\large S. K. Patra$^{1,3}$}
\affiliation{$^1$Institute of Physics, Bhubaneswar-751 005, India.}
\affiliation{$^2$ Saha Institute of Nuclear Physics, 1/AF,  Bidhannagar, Kolkata - 700064, India.}
\affiliation{$^3$ Homi Bhabha National Institute, Training School Complex, Anushakti Nagar, Mumbai 400085, India}
\maketitle

\section*{Introduction}
In the present scenario, nuclear physics and nuclear astrophysics are
well described within the self-consistent effective mean field models.
These effective theories are not only successful to describe finite
nuclei properties but also explain well the nuclear matter system at
supra-normal densities.  The  relativistic as well as non-relativistic
effective models are only the viable means to describe the  nuclear
systems over a wide range of masses. In recent decades, using the
relativistic and non-relativistic formalisms, a large number of nuclear
phenomena have been predicted near the nuclear drip-lines. As a results,
several experiments are planed in various laboratories to probe more
deeper side of the unknown nuclear territories, i.e., the regions of
neutron and proton drip-lines.

Here we present the results for newly proposed parameterizations IOPB-I
along with the recently published two parameter sets G3 \cite{G3} and
FSU2R\cite{FSU2R}.  We compare the behavior of the equation of states
(EOSs) of pure neutron matter (PNM) at sub-saturation densities for the
different parameterization considered.  The saturation properties for
the symmetry energy and slope of the symmetry energy of the modified
interactions are compared to recent experimental and theoretical
models. Finally, we calculate the mass-radius of the neutron stars for
the NL3, IOPB-I, FSU2R and G3 sets.

\begin{table}[t]
\caption{The obtained new parameter set IOPB-I along with NL3 \cite{NL3}, 
FSU2R \cite{FSU2R}, G3\cite{G3}  forces. The nucleon mass $M$ is 939.0 MeV. All the coupling constants are dimensionless, except $k_3$ which is in fm$^{-1}$.
The lower portion of the table indicates the nuclear matter properties such as
binding energy per nucleon $\mathcal{E}_{0}$(MeV), saturation density $\rho_{0}$(fm$^{-3}$), incompressibility coefficient for symmetric nuclear matter 
$K_{\infty}$(MeV), effective mass ratio $M^{*}/M$, symmetry energy $J$(MeV) 
and slope of the symmetry energy $L$(MeV).}
\scalebox{1.2}{
\begin{tabular}{cccccccccc}
\hline
\multicolumn{1}{c}{}
&&\multicolumn{1}{c}{NL3}
&&\multicolumn{1}{c}{IOPB-I}
&&\multicolumn{1}{c}{FSU2R}
&&\multicolumn{1}{c}{G3}\\
\hline
\hline
$m_{s}/M$  & & 0.541  && 0.533 && 0.529  &&    0.559  \\
$m_{\omega}/M$  & & 0.833  && 0.833 && 0.833  &&    0.832  \\
$m_{\rho}/M$  & & 0.812  && 0.812 && 0.812  &&    0.820  \\
$m_{\delta}/M$  & & 0.0  && 0.0 && 0.0  &&    1.043  \\
$g_{s}/4 \pi$  &&  0.813  &&  0.827 && 0.825 & &    0.782  \\
$g_{\omega}/4 \pi$  &&  1.024  &&  1.062 && 1.074  &&    0.923 \\
$g_{\rho}/4 \pi$  &&  0.712  && 0.885 &&  1.1433 &&   0.962   \\
$g_{\delta}/4 \pi$  &&  0.0  && 0.0 &&  0.0  &&    0.160  \\
$k_{3} $   &&  1.465  && 1.495 &&  1.261  &&  2.606  \\
$k_{4}$  &&  -5.688  && -2.932 &&  -0.644  &&  1.694   \\
$\zeta_{0}$  &&  0.0  && 3.103 &&  4.377 &&  1.010    \\
$\eta_{1}$  &&  0.0  && 0.0 &&  0.0  &&   0.424   \\
$\eta_{2}$  &&  0.0  && 0.0 &&  0.0  &&   0.114   \\
$\eta_{\rho}$  &&  0.0  && 0.0 &&  0.0  &&  0.645   \\
$\eta_{2\rho}$  &&  0.0  && 18.258 &&  56.275  &&   33.250   \\
$\alpha_{1}$  &&  0.0  && 0.0 &&  0.0  &&   2.000  \\
$\alpha_{2}$  &&  0.0  && 0.0 &&  0.0  &&  -1.468  \\
$f_\omega/4$  &&  0.0  &&  0.0 && 0.0  &&   0.220 \\
\hline
 &  &  &  &  & &&&\\
$\mathcal{E}_{0}$  &&  -16.29  && -16.10 &&  -16.28  &&    -16.02 \\
$\rho_{0}$  &&  0.148  && 0.149 &&  0.150  &&   0.148   \\
$K_{\infty}$  &&  271.5  && 222.6 &&  238.0  &&    243.9   \\
$M^*/M$  &&  0.595  && 0.593 &&  0.593  &&   0.699   \\
$J$  &&  37.40  && 33.30 &&  30.7  &&   31.8   \\
$L$  &&  118.6  && 63.6 &&  55.7  &&  49.3   \\
\hline
\end{tabular}}
\label{table1}
\end{table}

\section*{Results and Discussions}
\begin{figure}
\includegraphics[width=122mm]{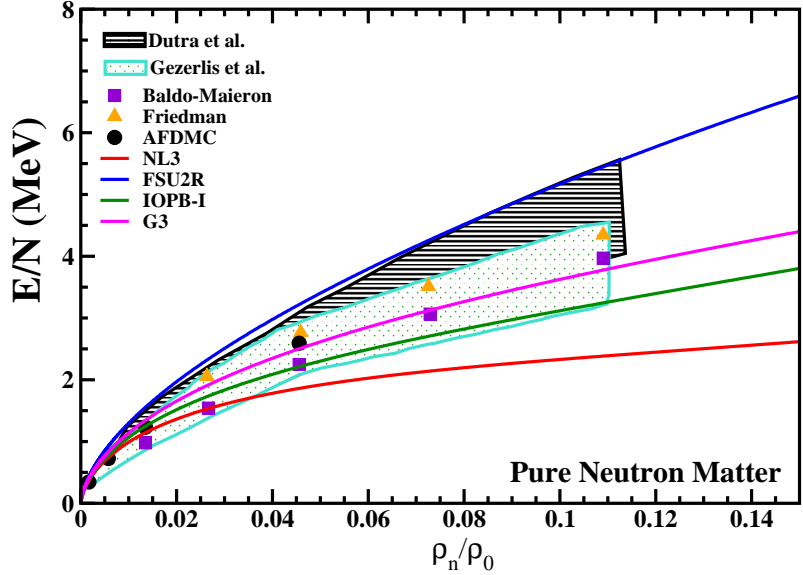}
\caption{\label{fig1} (Colour online) The binding energy per neutron as a function of
neutron density for the region of sub-saturation density.}  

\end{figure}

The IOPB-I set is obtained from the effective field theory motivated
relativistic mean field (ERMF) energy density functional, which
includes the contributions from $\delta$-meson to the lowest order
and the cross-coupling between $\omega$ and $\rho$-meson \cite{G3}.
The optimization of the energy density functional is performed for a given
set of fitting data using the simulated annealing method. The parameters
fitted with experimental data for the binding energies and  charge radii
for $^{16}$O, $^{40}$Ca, $^{48}$Ca, $^{68}$Ni, $^{90}$Zr, $^{100,132}$Sn,
and $^{208}$Pb nuclei with some constraints on the properties of the
nuclear matter at the saturation density. The newly developed IOPB-I
parameters along with other sets are displayed in Table \ref{table1}.
Fig. \ref{fig1}, displays the energy per neutron in pure neutron matter at
sub-saturation densities, which are encountered in finite nuclei and in
clusterization of nucleons. The results of NL3 set deviates significantly
and FSU2R, IOPB-I sets are marginally deviated from the experimental data
(shaded regions). However, parameter set G3 passes well through the
shaded region.

\begin{figure}
\includegraphics[width=122mm]{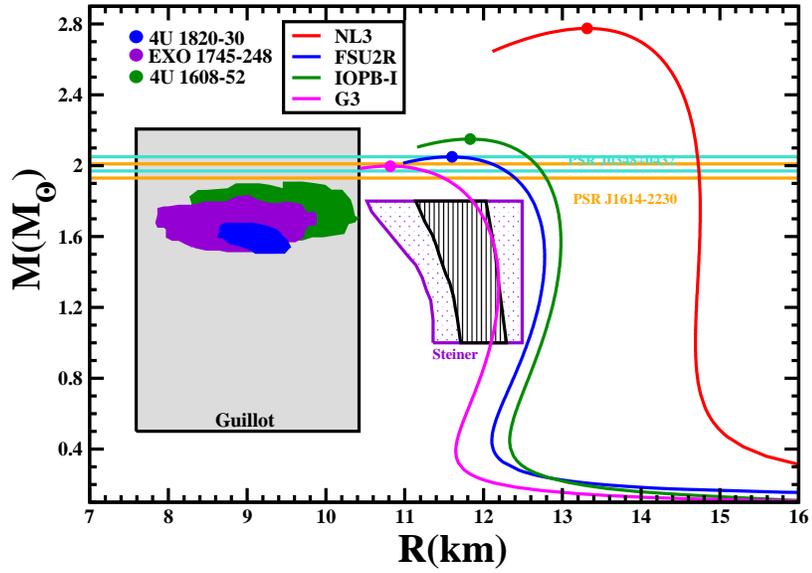}
\caption{\label{fig2} (Colour online) Mass versus radius of the neutron stars for various EOSs.
 }
\end{figure}

Finally, we use our parameter set to evaluate the mass and radius of
the static neutron star composed of neutrons, protons, electrons and
muons and the computed results are shown in Fig. \ref{fig2}.  From recent
observations, it is clearly illustrated that the maximum mass predicted by
any theoretical model should reach the limit $\sim$2.0$M_\odot$, which is
consistent with our present prediction from the G3 equation of state of
a nucleonic matter compact star with mass 1.997$M_\odot$ and radius 10.82
km. The IOPB-I set gives a larger and heavier NS with mass 2.15$M_\odot$
and radius 11.83 km comparable to the prediction of FSU2R set\cite{FSU2R}.

\section*{Summary and Conclusions}
In the present report, we have constructed a new parameter set termed as
IOPB-I for the ERMF model by fitting the experimental binding
energy and charge radius of eight spherical nuclei. In addition to
the finite nuclei data, we have also imposed some constraint on
the nuclear matter at saturation density. The neutron matter EOS at
sub-saturation densities for IOPB-I and G3 parameter sets show reasonable
improvement over other considered parameter sets. The maximum mass of
G3 parameter set is compatible with the measurements and the radius
of the neutron star with the canonical mass agree quite well with the
empirical values. However, IOPB-I set gives larger mass and radius
of the neutron star.

\end{document}